\documentclass[final,numberedheadings]{aipproc}
\layoutstyle{8x11single}

\newcommand{\mg}{\langle}
\newcommand{\md}{\rangle}
\newcommand{\vchi}{{v}}
\newcommand{\bk}{{\bf k}}
\newcommand{\bx}{{\bf x}}
\newcommand {\dphi}{{\delta\!\varphi}}
\newcommand{\chib}{{\bar \chi}}

\newcommand{\chis}{\chi_{_{\rm S}}}
\newcommand{\dchis}{\delta\chi_{_{\rm S}}}
\newcommand{\chisO}{\chi_{_{\rm S}}^{(0)}}
\newcommand{\dchisO}{\delta\chi_{_{\rm S}}^{(0)}}
\newcommand{\chisOb}{{\overline{\chi_{_{\rm S}}^{(0)}}}}

\newcommand{\mF}{{\cal F}}
\newcommand{\mT}{{\cal T}}
\newcommand{\vk}{{\bf k}}
\newcommand{\vx}{{\bf x}}
\newcommand{\dd}{{\rm d}}
\newcommand{\ii}{{\rm i}}

\begin{document}

\title{Models of inflation with primordial non-Gaussianities}

\classification{98.80.{-}
k ; 98.80.Cq ;  98.70.Vc}
\keywords{Cosmology, Early Universe, Inflation, Cosmic Microwave Background}

\author{Francis Bernardeau}{address={Service de Physique Th{\'e}orique,
         CEA/DSM/SPhT, Unit{\'e} de recherche associ{\'e}e au CNRS, CEA/Saclay
         91191 Gif-sur-Yvette c{\'e}dex}}
         
\author{Tristan Brunier}{address={Service de Physique Th{\'e}orique,
         CEA/DSM/SPhT, Unit{\'e} de recherche associ{\'e}e au CNRS, CEA/Saclay
         91191 Gif-sur-Yvette c{\'e}dex}}

\author{Jean--Philippe Uzan}{address={Institut d'Astrophysique de Paris, UMR7095 CNRS, Universit{\'e} Pierre \& Marie Curie - Paris, 98 bis bd Arago, 75014 Paris, France}}

\begin{abstract}
We present a class of models in which the primordial metric fluctuations do not necessarily obey Gaussian statistics. These models are realizations of mechanisms in which
non-Gaussianity is first generated by a light scalar field and
then transferred into curvature fluctuations during or at the end of inflation. 
For this class of models we present generic results for the probability
distribution functions of the metric perturbation at the end of
inflation. It is stressed that finite volume effects can induce non trivial effects that we
sketch.
\end{abstract}

\maketitle


\section{Introduction}

Cosmic Microwave Background (CMB) observations offer a precious window for the physics of the early Universe. 
And it is well known that inflation generically predicts Gaussian initial metric
fluctuations with an almost scale invariant power
spectrum~\cite{inflation}.  With the advent of large scale
structure and CMB surveys however it will be possible to test in details
the statistical properties of the initial conditions.  So far no
non-Gaussian signal has been detected in CMB data~\cite{cmb} but
the number of modes that can be probed is still small. In
large-scale structure surveys the number of independent modes that
are observed is large but the difficulty arises from the
non-linear gravitational dynamics~\cite{PTrevue} which can shadow
the primordial Non-Gaussianity (NG). Having at our disposal
models of inflation in which non-Gaussian adiabatic metric
fluctuations are generated can then serve as a guideline for
designing strategies for detecting primordial non-Gaussianities, in particular in the context of the Planck mission.

A number of propositions have been made which circumvent the usual bounds on primordial NG generic inflation (e.g. slow-roll single field inflation) predicts (see below).
This is the case in particular when primordial adiabatic fluctuations are subdominant compared to density fluctuations generated during the pre- or re-heating phases. Examples of such models are the curvaton model (which relies on assumptions on the reheating phase~\cite{curvaton}) or
recent propositions of multiple-field inflation which are based on relatively well understood phases of preheating~\cite{KolbVallinottoWands}.

%
%

\section{Generic inflation}
\label{GenInflation}

As mentioned in the introduction, we are interested in models that 
can produce sufficiently large non-Gaussianity. To be more precise, that 
should be the case for at least the range of modes that are observationally relevant,
i.e. that corresponds to the large scale structure scales. 

Let us start by considering a single field, $\varphi$, in slow-roll
inflation. The Klein-Gordon equation for its perturbation, $\dphi$, is of
the form
\begin{equation}\label{ov:1}
\ddot{\dphi}+3H\dot{\dphi}-\frac{\Delta}{a^2}\dphi=-V''\dphi
-V'''\frac{\dphi^2}{2}+\ldots
\end{equation}
During the slow-roll regime, $\vert\dot H\vert\ll H^2$ and
$\ddot\varphi\ll3H\dot\varphi$ so that\footnote{We have set $M_4^{-2}\equiv8\pi G$.}
\begin{equation}\label{ov:2}
 3H^2M_4^2\simeq V,\qquad
 H\dot\varphi\simeq -V'
\end{equation}
and the slow-roll conditions can be expressed as $\varepsilon\ll1$
and $|\eta|\ll1$ with
\begin{equation}\label{ov:3}
  \varepsilon\equiv M_4\frac{V'}{V},\qquad
  \eta\equiv M_4^2\frac{V''}{V}.
\end{equation}
In order for the fluctuations of the scalar field to be large
enough, one needs the mass of the field to be much smaller than
$H$, in which case one gets $\left<\delta\varphi\right>\sim H$. Furthermore to
get the correct amplitude for the primordial fluctuations, one
should have $V^{3/2}/(V'M_4^3)\sim10^{-5}$ so that
\begin{equation}\label{ov:4}
 H\sim10^{-5}\varepsilon M_4.
\end{equation}
The range of wavelengths that correspond to the observed  large scale
structure exits the Hubble radius over the number of $e$-foldings
\begin{equation}\label{ov:5}
  N_\lambda=H\Delta t.
\end{equation}
During that period, the slow-roll parameter $\eta$ has varied over the range
$\Delta\eta\sim(\xi-\eta\varepsilon)\Delta\varphi/M_4$, with
$\xi\equiv M_4^3(V'''/V)$. From Eq.~(\ref{ov:4}), one deduces that
\begin{equation}\label{ov:5bis}
 \Delta\varphi/M_4\sim-\varepsilon H\Delta t.
\end{equation}
Now, if the quadratic term in the r.h.s. of Eq~(\ref{ov:1}) is
dominant over the linear term then $V''/V'''<\dphi\sim H$
and, using Eq~(\ref{ov:4}), one deduces that $\Delta\eta>10^5\eta
N_\lambda$. This will induce a rapid breakdown of the slow-roll
inflation. This can be understood from the fact that the potential
has to be both flat enough for the fluctuations to develop and
steep enough for the non-linear terms not to be negligible\footnote{A way
round to this argument is to consider potential with a sharp
feature~\cite{sharpfeature} but in that case the non-Gaussianity is
associated with a departure from scale invariance and is located
on a
very small band of wavelengths.}

\section{Multiple field inflation}

\begin{figure}
\includegraphics[width=9cm]{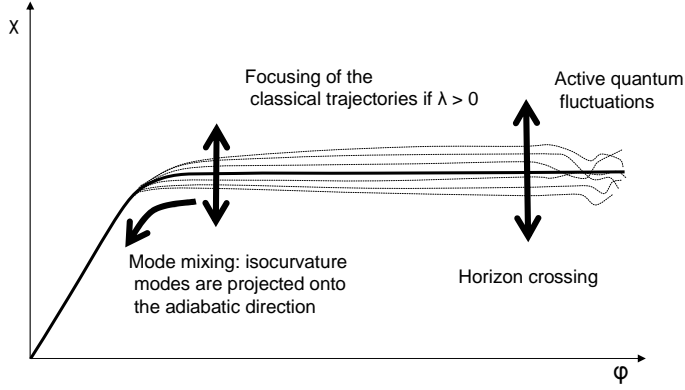}
\caption{The trajectories of the fields in the plane ($\phi,\chi$),
  once smoothed on a scale $R$. Before horizon crossing ($R<H$), the
  trajectories behave quantumly because the quantum fluctuations are
  active up to scale $H$ and are not smoothed out. After the horizon crossing
  ($R>H$), the trajectories can be treated as classical
  trajectories. Note that this transition happens at different time
  for different values of $R$. The bundle of classical trajectories
  then evolves in the two dimensional potential and its cross section
  evolves with time. In the case sketched here, because of the bending of the potential valley, the
  isocurvature modes induce metric fluctuations because of 
  differences in length of the trajectories. The net result is a transfer of modes between isocurvature and adiabatic directions.}
\label{NGTrajFocus}
\end{figure}

The previous analysis shows that it is not possible to get significant NG effects over the observable wavelengths with single field inflation. To obtain such effects, extra fields are necessary.
A class of multiple field inflationary model corresponds to a Lagrangian where $N$ scalar fields
are minimally coupled to the metric, e.g.,
\begin{equation}\label{eq:lagrangien}
{\cal L}=-\frac{R}{16\pi
G}+\frac{1}{2}\sum_{j=1}^N\partial_\mu\varphi_j
\partial^\mu\varphi_j - V(\varphi_1,\ldots,\varphi_N),
\end{equation}
where $R$ is the Ricci scalar. 
It follows that the background Einstein equations take the form,
\begin{eqnarray}
H^2=\frac{4\pi G}{3}\left(\sum_{j=1}^N\dot\varphi_j^2 +2V
\right),\qquad\dot H=-4\pi G\sum_{j=1}^N\dot\varphi_j^2,\qquad\ddot\varphi_j+3H\dot\varphi_j=-V_j,\label{eq:back}
\end{eqnarray}
where a dot refers to a derivation with respect to the cosmic
time, $t$, $H\equiv\dot a/a$ and $V_{j}$ refers to a derivation
with respect to $\varphi_j$. 

Then, let us consider the simplest case in which we have only two scalar
fields $\varphi_1$ and $\varphi_2$. We decompose these two fields into the genuine inflationary direction $\dphi$ and a transverse direction $\chi$, taking advantage of
the direction defined by the inflationary direction,
\begin{equation}\label{eq:rotation}
 \pmatrix{\dot \dphi \cr \dot \chi}
 =
 {\cal M}(\theta)
 \pmatrix{\dot\varphi_1 \cr \varphi_2},\qquad
 {\cal M}(\theta)
 \equiv
   \pmatrix{
     \cos\theta & \sin\theta \cr
     -\sin\theta & \cos\theta},
\end{equation}
the angle $\theta$ being defined by
\begin{equation}\label{eq:def-theta}
\cos\theta\equiv\frac{\dot\varphi_1}{\sqrt{\dot\varphi_1^2+\dot\varphi_2^2}},
\qquad
\sin\theta\equiv\frac{\dot\varphi_2}{\sqrt{\dot\varphi_1^2+\dot\varphi_2^2}}.
\end{equation}
{}From these definitions it follows that $\chi$ is stationary along the classical inflationary trajectory. To be more precise, the Klein-Gordon equations imply that,
\begin{eqnarray}
\ddot\dphi+3H\dot\dphi-\frac{\Delta \dphi}{a^2}&=&-\frac{\partial^2 V}{\partial \dphi^2}\dphi-\frac{\partial^2 V}{\partial \dphi\partial\chi}\chi,\label{KGdphi}\\
\ddot\chi+3H\dot\chi-\frac{\Delta \chi}{a^2}&=&...-\frac{\partial^3 V}{\partial \chi^3}\frac{\chi^2}{2}-\frac{\partial^4 V}{\partial\chi^4}\frac{\chi^3}{3!}\label{KGchi}.
\end{eqnarray}
This set of equation clearly opens the way to new phenomena. Indeed, nothing prevents the fluctuations in the $\chi$ directions to develop significant non-Gaussian properties. This would be the case for instance for generic quartic potentials. Those field fluctuations can appear as metric fluctuations if, at some stage before the end of the inflationary period, the $\chi$ fluctuations can be transferred into the adiabatic fluctuation. According to Eq. (\ref{KGdphi}), this is possible provided ${\partial^2 V}/({\partial \dphi\partial\chi})$ is not zero. Geometrically speaking, it corresponds to a bent in the inflationary trajectory\footnote{Note that this bent can also be encoutered during the preheating phase. An example of such a model in presented in~\cite{BKU}.}~\cite{bartolo,ub02}. This idea is sketched on diagram \ref{NGTrajFocus}. 



\subsection{A simple working example: extended hybrid inflation}

Although generic predictions can be made for such a class of models, it is to be stressed that such a scenario does not require elaborate or exotic field theory constructions. Let's for instance consider the following multiple-field potential presented in \cite{BUPRL},
\begin{eqnarray}\label{threefieldmodel}
 V(\phi,\chi,\sigma)&=&\frac{1}{2}m^2\,\phi^2
        +\frac{\lambda}{4!}\chi^4+
        \frac{\mu}{2}\left(\sigma^2-\sigma_0^2\right)^2
        +\frac{g}{2}\sigma^2\left(\phi\cos\alpha+\chi\sin\alpha\right)^2.
\end{eqnarray}
It is an extension of the hybrid model with one extra field. Here, 
one field, $\phi$, is the inflaton; the second field is a light
scalar, $\chi$, with quartic coupling $0<\lambda\le 1$ and the third
field, $\sigma$, is coupled to the two others so that the end of
inflation is triggered when $\sigma$ undergoes a phase transition and $\sigma_0$ is the final vev of $\sigma$. One extra parameter $\alpha$ describes the
mixing angle between $\phi$ and $\chi$ in their coupling to
$\sigma$. This parameter will determine the ratio between the initial isocurvature and adiabatic modes in the final metric fluctuations. Note that a quartic interaction has been introduced since this is the only one that does not require further fine tuned parameter (any other polynomial type interaction would require dimensioned parameter tuned to $H$, see \cite{ub02}).

In this model, as for hybrid inflation, the inflationary stage ends when the effective mass of $\sigma$ vanishes, that is when
\begin{equation}\label{effectivemass}
  g\left(\phi\cos\alpha+\chi\sin\alpha\right)^2-2\mu\sigma_0^2=0.
\end{equation}
The value of $\phi$ at the end of inflation is therefore
\begin{equation}\label{phiend}
  \phi_{\rm
  end}\equiv\frac{\pm\sqrt{2\mu/g}\,\sigma_0-\chi\sin\alpha}{\cos\alpha}.
\end{equation}
For $\phi>\phi_{\rm end}$, $\sigma=0$ and the two fields evolve
independently: $\phi$ drives the inflation while $\chi$ develops
non-Gaussianity. The amount of non-Gaussianity of $\chi$ then
depends only on $\lambda$ and on the total number of e-foldings
between horizon crossing and the end of inflation.

\subsection{Test quantum scalar field in de Sitter space}
\label{tqsf}

The non-Gaussian properties of the $\chi$ field during the inflationary phase are those developed by a self interacting scalar field in an expanding universe. The full resolution of this problem in quite involved. It can be addressed with a Perturbation Theory approach that can be done either at a quantum level or, once the horizon has been crossed, at a classical level (assuming the scalar field behaves like a classical stochastic field).


In \cite{BBU} we present the result of the computation of the high order correlation function of a test scalar field in a de Sitter or quasi de Sitter background. The calculation is based on an expansion of the field evolution from the free field solution as summarized in the following. For a minimally coupled free quantum field, the solution
can indeed be decomposed in plane waves as,
\begin{equation}\label{eq:quantdec}
 \widehat \vchi_0(\bx,\eta)=\int\dd^3\bk
 \left[\vchi_0(k,\eta)\widehat b_\bk\hbox{e}^{i\bk\cdot\bx}
 +\vchi^*_0(k,\eta)\widehat b_\bk^\dag\hbox{e}^{-i\bk\cdot\bx}\right],
\end{equation}
where we have introduced $\widehat \vchi\equiv a\,\widehat\chi$, a
hat referring to a quantum operator and $b_\bk$ and $b_\bk^\dag$ and the annihilation and creation operators for a particle of momentum $\vk$ in a Bunch-Davies vacuum. In the massless limit, the free field solution is
\begin{equation}
\vchi_0(k,\eta)=\left(1-\frac{\ii}{k\eta}\right)\frac{\hbox{e}^{-\ii
k\eta}}{\sqrt{2k}}.
\end{equation}
One can then express
perturbatively the $N$-point correlation functions of the interacting
field, $\chi$, in terms of those of the free scalar field. The equal
time correlators are expectation values of
product of field operators for the current time vacuum state. Such
computations can be performed following general principles of quantum
field calculations~\cite{maldacena}. The simplest formulation
is to apply the evolution operator $U(\eta_0,\eta)$ backward in time
to transform the interacting field vacuum into the free field vacuum
at an arbitrarily early time $\eta_0$ so that,
\begin{equation}\label{CumulantGeneral}
  \mg \vchi_{\vk_1}\ldots\vchi_{\vk_n}\md\equiv
  \mg 0\vert U^{-1}(\eta_0,\eta)\ \vchi_{\vk_1}\ldots\vchi_{\vk_n}\ U(\eta_0,\eta)\vert
  0\md
\end{equation}
where $\vert 0 \md$ is here the \emph{free field} vacuum~\footnote{It
is to be noted that these calculations do not correspond to those of
diffusion amplitudes of some interaction processes in a de Sitter
space (see \cite{Weinberg} for a comprehensive presentation of those calculations.). When one tries to do these latter calculations with a path
integral formulations, mathematical divergences are encountered as it
has been stressed in~\cite{tsamis,ub02}. With that respect de Sitter
space strongly differs from Minkowski space-time.}.

The evolution operator $U$ can be written in terms of the interaction
Hamiltonian, $H_I$, as
\begin{equation}\label{Uoperator}
U(\eta_0,\eta)=\mT\exp{\left(-\ii \int_{\eta_0}^{\eta}\dd\eta'\ H_I(\eta')\right)}
\end{equation}
where $\mT$ is the time ordered product operator\footnote{Then the inverse operator for $U$ reads, $U^{-1}(\eta_0,\eta) = \bar{\mT} \exp\left( +\ii
\int_{\eta_0}^{\eta} {\rm d} \eta' \; H_I(\eta') \right)$ where $\bar{\mT}$ is the inverse $\mT$ product.}.
Eventually the connected part of the above ensemble average at a time
$\eta$ reads,
\begin{equation}\label{Cumulant}
\mg
  \vchi_{\vk_1}\ldots\vchi_{\vk_n}\md_c=-\ii\int_{\eta_0}^{\eta}\dd\eta'\
  \mg 0\vert
  \left[\vchi_{\vk_1}\ldots\vchi_{\vk_n},H_I(\eta')\right]\vert 0\md,
\end{equation}
where the brackets stand for the commutator.

The result for the four-point function in case of a quartic interaction (with parameter $\lambda$) can finally be written as,
\begin{equation}
  \mg \chi_{\vk_1}\ldots\chi_{\vk_4}\md_c =
      \delta_{\rm Dirac}\left(\sum \vk_i\right)P_4(k_1,k_2,k_3,k_4),\qquad P_4(k_1,k_2,k_3,k_4)=\nu_3(\{k_i\})\sum_{i} \prod_{j \ne i}
  \frac{H^2}{2 k_j^3},\label{p4sh}
\end{equation}
where the vertex value, $\nu_3$, reads
\begin{equation}\label{vertexvalue}
  \nu_3(\{k_i\})=\frac{\lambda}{3 H^2}\left[\zeta(\eta,k_i)
  +\log\left(-\eta\sum k_i\right)\right].
\end{equation}
In this expression $\zeta(\eta,k_i)$ can be expressed\footnote{note that $\zeta$ differs from the one introduced in \cite{BBU}.} in terms of the four wavelength $k_{i}$. It is finite for any values of $\eta$ and $k_{i}$ (in particular in the super-Hubble limit).

\begin{figure}
\includegraphics[width=9cm]{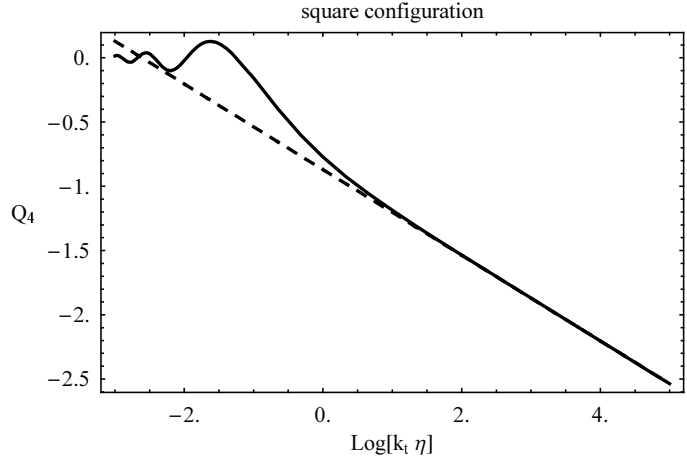}
\caption{Behaviour of the function $Q_4$ as of function of time.  The
transition to the superhorizon behavior (dashed line) is shown. The
function $Q_4$ is shown here for a ''square" configuration
($k_1=k_2=k_3=k_4$) as a function of $k_t\eta=\sum k_i\eta$.}
\label{Q4square}
\end{figure}

When the term $\log\left(-\eta\sum k_i\right)$ is large (and
negative), that is when the number of $e$-foldings, $N_e$, between the
time of horizon crossing for the modes we are interested in and the
end of inflation is large, the vertex value is simply given by,
$$
\nu_3(\{k_i\})=-{\lambda\,N_e}/{(3\,H^2)},
$$
which corresponds exactly to what a classical stochastic approach would give~\cite{ub02}.
On Fig. \ref{Q4square} one can then appreciate the transition from a regime which is dominated by quantum fluctuations to a regime following a classical evolution.  In this figure, what is plotted is the reduced four-point correlation function, $Q_4$, defined as,
\begin{equation}\label{q4}
  Q_4(\{k_i\})=\frac{P_4(k_1,k_2,k_3,k_4)}{P_2(k_1)P_2(k_2)P_2(k_3)+{\rm
  sym.}}
\end{equation}
where $P_{2}(k)$ is the free field power spectrum.

These results illustrate one quantitative aspects of the NG properties that self-coupled scalar fields can develop, namely high order correlation functions, whose computation is given here at tree order. 
These calculations are on solid ground. They do not give however a complete prescription for the description of the statistical properties of the field.

\subsection{PDF and finite volume effects, a classical stochastic approach}

The use of a classical perturbative approach can provide us with a complete description of the expected properties of the resulting field properties. In particular it is possible to infer the whole one point probability distribution function (PDF) of the field value. 

The approach detailed in~\cite{ub02} and~\cite{FiniteVol} is based on the expansion
the filtered field in terms of the coupling constant as
\begin{equation}\label{chiexp}
  \chis(\eta)=\chis^{(0)}(\eta)+\chis^{(1)}(\eta)+\dots.
\end{equation}
The subscript ${\rm S}$ stands for the fact that the field has been convolved with a given smoothing window function in such a way that only large
enough scales are taken into account.
 Then 
$\chis^{(0)}$ represents the value of the filtered field when the
self-interacting term in the potential is dropped. In the slow-roll regime the
field $\chis$ follows the Klein-Gordon equation~(\ref{KGchi}). Then $\chis^{(0)}$ is simply constant during the slow roll evolution. It represents the initial conditions for the non-linear field evolution.  
At first order term in $\lambda$, $\chis\equiv{\chis}^{(1)}$, evolves
according to
\begin{equation}\label{dsl1evol}
  3H\dot{\chis}^{(1)}=
  -\frac{\lambda}{3!}\left[\chis^{(0)}\right]^3
  \end{equation}
for a quartic potential. In this approach the treatment of the filtering of the r.h.s. of
this equation is very crude. This is the price to pay for using this approach. The results we are going to find can
anyway be checked against the more rigorous calculations based on the
computation shape of the tri-spectrum presented in the previous subsection. 

Resolving Eq. (\ref{dsl1evol}), leads to the expression of ${\chis}^{(1)}$ as a function of $\chis^{(0)}$. 
$\chi_{\rm S}$ is actually built out from modes whose wavelengths are
larger than the smoothing scale. During inflation, $\chi_{\rm S}$ can be
viewed as a classical random field as soon as the smoothing scale has
crossed the comoving horizon. Today, $\chi_{\rm S}$ is made of a the
superposition of modes that are super-horizon and of modes which are now
within our horizon. It is important to understand that, although the
ensemble average of $\chi_{\rm S}$ actually vanishes, its geometrical mean
at the survey size, as given by observations, is expected to be non-zero
(since at least super-horizon modes are out of reach).

The value of $\chi_{\rm S}$ at our horizon scale, say $\chisOb$, appears therefore as a new free parameter. Its value is determined by the peculiar realization of the inflationary path our Universe has followed. In our model case $\chis^{(0)}$ is bounded ($\lambda>0$) and its range can be estimated through a Fokker-Planck equation~\cite{FokkerPlanck,FiniteVol}. Phenomenologically it means that one should decompose $\chis^{(0)}$ as,
\begin{equation}
\chis^{(0)}=\dchisO+\chisOb.
\end{equation}
where $\dchisO$ is then Gaussian distributed with zero average and given variance $\sigma_{\delta}$. Similarly,
\begin{equation}
\chis=\dchis+\chib
\end{equation}
where $\chib$ corresponds to the averaged value of $\chi$ at our horizon size. 

The equation of evolution~(\ref{dsl1evol}) can be solved to get
\begin{equation}\label{dsl1sol}
  \chis^{(1)}(t)=-\lambda(t-t_{_{\rm H}})\frac{\left(\chis^{(0)}\right)^3}{18\,H},
\end{equation}
which also reads
\begin{equation}\label{dsl1sol2}
  \chis^{(1)}(t)=-\frac{\lambda N_e}{18\,H^2}\left(\chis^{(0)}\right)^3,
\end{equation}
$N_e$ being the number of $e$-foldings between $t_H$ and the end of
inflation.

These results imply that
\begin{equation}\label{chisbar}
  \chib\approx \chisOb-\frac{\lambda N_e}{18\,H^2}\left[\left(\chisOb\right)^3+
  3\,\chisOb\,\sigma_\delta^2
  \right]
\end{equation}
which explicitly shows that $\chib$ and $\chisOb$ are equal at
leading order in $\lambda$. It also gives
\begin{equation}\label{dchis}
  \dchis=\dchisO-\frac{\lambda N_e}{18\,H^2}\left\lbrace
  \left(\dchisO\right)^3+
  3\left[\left(\dchisO\right)^2-\sigma_{\delta}^2\right]\chib +
  3\dchisO\chib^2 \right\rbrace.
\end{equation}
It is straightforward to see that $\dchis$ has acquired a non zero
third order moment,
\begin{equation}\label{skewnessdchis}
  \overline{\dchis^3}=-\frac{\lambda
  N_e}{H^2}\chib\sigma_{\delta}^4
\end{equation}
at leading order in $\lambda$. This is a finite volume effect in
the sense that it exists for a fixed (not ensemble averaged) value
of $\chib$. This effect cannot a priori be neglected. It can be shown that $\chib$ should be of
the order of $H/\lambda^{1/4}$, which implies  that the reduced
skewness of $\dchis$,
$\overline\dchis^3/(\overline\dchis^2)^{3/2}\sim
\lambda^{3/4}N_e\sigma_{\delta}/H$ is significant as soon as
$\lambda^{3/4}N_e$ approaches unity.

An attentive reader cannot but notice that the evolution equation for $\chis$ can in fact be solved in
\begin{equation}\label{chissol}
\chis=\frac{\chisO}{\sqrt{1-\frac{\lambda\,N_e}{9H^2}\left(\chisO\right)^2}}
\end{equation}
and the distribution of $\chis$ can then be infered from that of
$\chisO$ assuming the latter is Gaussian distributed with a
non-zero mean value~\footnote{As noted in Ref.~\cite{ub02}, such a
simple variable change implicitly incorporates ``loop order"
effects that, because of sub-Hubble physics, are not necessarily
correctly estimated. In that paper we developed a more elaborated
method which allows the reconstruction of the PDF from the only
tree order contributions of each cumulant. As the two approaches
eventually give the same qualitative results we restrict here our
analysis to the simplest method.}.

\begin{figure}
\includegraphics[width=9cm]{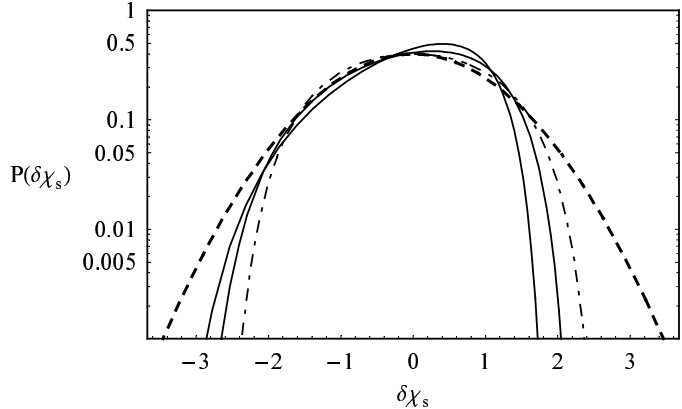}
 \caption{PDF of $\delta\chis=\delta\chis-\chib$ for different
values of $\chib$. The dashed line corresponds to a Gaussian
distribution; the dot-dashed line to the deformed distribution of
$\dchis$ when $\lambda N_e/H^2=1$ and $\chib=0$ and the solid
lines when the latter equals $0.5$ and $1$.} \label{fvepdf}
\end{figure}

In figure~\ref{fvepdf}, we present the deformation of the PDF of
$\delta\chis$ while $\chib$ is varied. As expected, it shows that
when $\chib$ is not zero, the PDF gets skewed in a way that can be
easily understood: when $\chib$ is positive it gets more difficult
to have excursion towards larger value of $\chis$, but easier to
roll down to smaller values. It is as if the field $\chi$ was
actually evolving in the potential $\lambda(\chi+\chib)/4!$. 

What should then be expected for observations? In this family of models, the surviving couplings in the metric, are expected to be, to a good approximation, equivalent to those induced by the superposition of two stochastically independent fields, a Gaussian one and one obtained by a non-linear transform of a Gaussian field with the same spectrum, e.g. Eq. (\ref{chissol}). In other words the local (Bardeen) potential would read,
\begin{equation}
\Phi(\vx)=\cos \alpha\ \Phi_{1}(\vx)+ \sin \alpha\ \mF[\Phi_{2}(\vx)]
\end{equation}
where ratio between the initial adiabatic and isocurvature fluctuations is described by a mixing angle $\alpha$ and where the function $\mF$ takes into account the self coupling of the field $\chi$ that gave rise to this part of the metric fluctuations. This description is the starting point of investigations for the CMB properties presented in~\cite{lastpaper}.


\section{Conclusions}

We have shown here that it is possible
to build explicit models of inflation where primordial metric fluctuations can develop significant 
non-Gaussian features. The class of models we have identified induce NG properties that can be described by a few identified parameters. Two of them are of fundamental origin, the amplitude of the  quartic coupling and the mixing angle; a third is due to finite volume effects and its value depends on the peculiar realization of the universe we live in, the transverse field value at Hubble size. Unlike most models proposed in the literature, they provide us with well defined properties of the metric fluctuations that can serve as a starting point for further investigations on
its various observational aspects, both for CMB and large scale
structure surveys~\cite{lastpaper}. In particular, the models do not require to deal with the more
intricate second order dynamics.


\begin{theacknowledgments}
 The writing of the series of papers whose content is partly presented here has benefited from numerous discussions with many colleagues. We more particularly thank Simon Prunet and Lev Kofman for many fruitful discussions concerning observational or theoretical issues.
The research of FB is supported in part by the ACI Jeunes Chercheurs 2068.
 
\end{theacknowledgments}

\end{document}